\documentclass[amsmath,amssymb,aps,nofootinbib,nolongbibliography,superscriptaddress]{revtex4-2}

\usepackage{bm}
\usepackage{dcolumn}
\usepackage{graphicx}
\usepackage{multirow}
\usepackage{color}
\usepackage{mathtools}
\usepackage[T1]{fontenc}
\usepackage{lmodern}

\begin{document}

\title{Complementarity between atmospheric and super-beam neutrinos at ESSnuSB}

\author{J.~Aguilar}
\affiliation{Consorcio ESS-bilbao, Parque Cient\'{i}fico y Tecnol\'{o}gico de Bizkaia, Laida Bidea, Edificio 207-B, 48160 Derio, Bizkaia, Spain}
\author{M.~Anastasopoulos}
\affiliation{European Spallation Source, Box 176, SE-221 00 Lund, Sweden}
\author{D.~Bar\u{c}ot}
\affiliation{Center of Excellence for Advanced Materials and Sensing Devices, Ruđer Bo\v{s}kovi\'c Institute, 10000 Zagreb, Croatia}
\author{E.~Baussan}
\affiliation{IPHC, Universit\'{e} de Strasbourg, CNRS/IN2P3, Strasbourg, France}
\author{A.K.~Bhattacharyya}
\affiliation{European Spallation Source, Box 176, SE-221 00 Lund, Sweden}
\author{A.~Bignami}
\affiliation{European Spallation Source, Box 176, SE-221 00 Lund, Sweden}
\author{M.~Blennow}
\affiliation{Department of Physics, School of Engineering Sciences, KTH Royal Institute of Technology, Roslagstullsbacken 21, 106 91 Stockholm, Sweden}
\affiliation{The Oskar Klein Centre, AlbaNova University Center, Roslagstullsbacken 21, 106 91 Stockholm, Sweden}
\author{M.~Bogomilov}
\affiliation{Sofia University St. Kliment Ohridski, Faculty of Physics, 1164 Sofia, Bulgaria}
\author{B.~Bolling}
\affiliation{European Spallation Source, Box 176, SE-221 00 Lund, Sweden}
\author{E.~Bouquerel}
\affiliation{IPHC, Universit\'{e} de Strasbourg, CNRS/IN2P3, Strasbourg, France}
\author{F.~Bramati}
\affiliation{University of Milano-Bicocca and INFN Sez. di Milano-Bicocca, 20126 Milano, Italy}
\author{A.~Branca}
\affiliation{University of Milano-Bicocca and INFN Sez. di Milano-Bicocca, 20126 Milano, Italy}
\author{G.~Brunetti}
\affiliation{University of Milano-Bicocca and INFN Sez. di Milano-Bicocca, 20126 Milano, Italy}
\author{I.~Bustinduy}
\affiliation{Consorcio ESS-bilbao, Parque Cient\'{i}fico y Tecnol\'{o}gico de Bizkaia, Laida Bidea, Edificio 207-B, 48160 Derio, Bizkaia, Spain}
\author{C.J.~Carlile}
\affiliation{Department of Physics, Lund University, P.O Box 118, 221 00 Lund, Sweden}
\author{J.~Cederkall}
\affiliation{Department of Physics, Lund University, P.O Box 118, 221 00 Lund, Sweden}
\author{T.~W.~Choi}
\affiliation{Department of Physics and Astronomy, FREIA Division, Uppsala University, P.O. Box 516, 751 20 Uppsala, Sweden}
\author{S.~Choubey}
\email{Corresponding authors: S. Choubey (choubey@kth.se), T. Ohlsson (tohlsson@kth.se) and S. Vihonen (vihonen@kth.se)}
\affiliation{Department of Physics, School of Engineering Sciences, KTH Royal Institute of Technology, Roslagstullsbacken 21, 106 91 Stockholm, Sweden}
\affiliation{The Oskar Klein Centre, AlbaNova University Center, Roslagstullsbacken 21, 106 91 Stockholm, Sweden}
\author{P.~Christiansen}
\affiliation{Department of Physics, Lund University, P.O Box 118, 221 00 Lund, Sweden}
\author{M.~Collins}
\affiliation{Faculty of Engineering, Lund University, P.O Box 118, 221 00 Lund, Sweden}
\affiliation{European Spallation Source, Box 176, SE-221 00 Lund, Sweden}
\author{E.~Cristaldo Morales}
\affiliation{University of Milano-Bicocca and INFN Sez. di Milano-Bicocca, 20126 Milano, Italy}
\author{P.~Cupia\l}
\affiliation{AGH University of Krakow, al. A. Mickiewicza 30, 30-059 Krakow, Poland }
\author{D.~D'Ago}
\affiliation{INFN Sez. di Padova, Padova, Italy}
\author{H.~Danared}
\affiliation{European Spallation Source, Box 176, SE-221 00 Lund, Sweden}
\author{J.~P.~A.~M.~de~Andr\'{e}}
\author{M.~Dracos}
\affiliation{IPHC, Universit\'{e} de Strasbourg, CNRS/IN2P3, Strasbourg, France}
\author{I.~Efthymiopoulos}
\affiliation{CERN, 1211 Geneva 23, Switzerland}
\author{T.~Ekel\"{o}f}
\affiliation{Department of Physics and Astronomy, FREIA Division, Uppsala University, P.O. Box 516, 751 20 Uppsala, Sweden}
\author{M.~Eshraqi}
\affiliation{European Spallation Source, Box 176, SE-221 00 Lund, Sweden}
\author{G.~Fanourakis}
\affiliation{Institute of Nuclear and Particle Physics, NCSR Demokritos, Neapoleos 27, 15341 Agia Paraskevi, Greece}
\author{A.~Farricker}
\affiliation{Cockroft Institute (A36), Liverpool University, Warrington WA4 4AD, UK}
\author{E.~Fasoula}
\affiliation{Department of Physics, Aristotle University of Thessaloniki, Thessaloniki, Greece}
\affiliation{Center for Interdisciplinary Research and Innovation (CIRI-AUTH), Thessaloniki, Greece}
\author{T.~Fukuda}
\affiliation{Institute for Advanced Research, Nagoya University, Nagoya 464–8601, Japan}
\author{N.~Gazis}
\affiliation{European Spallation Source, Box 176, SE-221 00 Lund, Sweden}
\author{Th.~Geralis}
\affiliation{Institute of Nuclear and Particle Physics, NCSR Demokritos, Neapoleos 27, 15341 Agia Paraskevi, Greece}
\author{M.~Ghosh}
\affiliation{Center of Excellence for Advanced Materials and Sensing Devices, Ruđer Bo\v{s}kovi\'c Institute, 10000 Zagreb, Croatia}
\author{A.~Giarnetti}
\affiliation{Dipartimento di Matematica e Fisica, Universit\'a di Roma Tre, Via della Vasca Navale 84, 00146 Rome, Italy}
\author{G.~Gokbulut}
\affiliation{University of Cukurova, Faculty of Science and Letters, Department of Physics, 01330 Adana, Turkey}
\affiliation{Department of Physics and Astronomy, Ghent University, Proeftuinstraat 86, B-9000 Ghent, Belgium}
\author{C.~Hagner}
\affiliation{Institute for Experimental Physics, Hamburg University, 22761 Hamburg, Germany}
\author{L.~Hali\'c}
\affiliation{Center of Excellence for Advanced Materials and Sensing Devices, Ruđer Bo\v{s}kovi\'c Institute, 10000 Zagreb, Croatia}
\author{M.~Hooft}
\affiliation{Department of Physics and Astronomy, Ghent University, Proeftuinstraat 86, B-9000 Ghent, Belgium}
\author{K.~E.~Iversen}
\affiliation{Department of Physics, Lund University, P.O Box 118, 221 00 Lund, Sweden}
\author{N.~Jachowicz}
\affiliation{Department of Physics and Astronomy, Ghent University, Proeftuinstraat 86, B-9000 Ghent, Belgium}
\author{M.~Jenssen}
\affiliation{European Spallation Source, Box 176, SE-221 00 Lund, Sweden}
\author{R.~Johansson}
\affiliation{European Spallation Source, Box 176, SE-221 00 Lund, Sweden}
\author{E.~Kasimi}
\affiliation{Department of Physics, Aristotle University of Thessaloniki, Thessaloniki, Greece}
\affiliation{Center for Interdisciplinary Research and Innovation (CIRI-AUTH), Thessaloniki, Greece}
\author{A.~Kayis Topaksu}
\affiliation{University of Cukurova, Faculty of Science and Letters, Department of Physics, 01330 Adana, Turkey}
\author{B.~Kildetoft}
\affiliation{European Spallation Source, Box 176, SE-221 00 Lund, Sweden}
\author{K.~Kordas}
\affiliation{Department of Physics, Aristotle University of Thessaloniki, Thessaloniki, Greece}
\affiliation{Center for Interdisciplinary Research and Innovation (CIRI-AUTH), Thessaloniki, Greece}
\author{B.~Kova\u{c}}
\affiliation{Center of Excellence for Advanced Materials and Sensing Devices, Ruđer Bo\v{s}kovi\'c Institute, 10000 Zagreb, Croatia}
\author{A.~Leisos}
\affiliation{Physics Laboratory, School of Science and Technology, Hellenic Open University, 26335, Patras, Greece }
\author{A.~Longhin}
\affiliation{Department of Physics and Astronomy "G. Galilei", University of Padova and INFN Sezione di Padova, Italy}
\author{C.~Maiano}
\affiliation{European Spallation Source, Box 176, SE-221 00 Lund, Sweden}
\author{S.~Marangoni}
\affiliation{University of Milano-Bicocca and INFN Sez. di Milano-Bicocca, 20126 Milano, Italy}
\author{J.~G.~Marcos}
\affiliation{Department of Physics and Astronomy, Ghent University, Proeftuinstraat 86, B-9000 Ghent, Belgium}
\author{C.~Marrelli}
\affiliation{CERN, 1211 Geneva 23, Switzerland}
\author{D.~Meloni}
\affiliation{Dipartimento di Matematica e Fisica, Universit\'a di Roma Tre, Via della Vasca Navale 84, 00146 Rome, Italy}
\author{M.~Mezzetto}
\affiliation{INFN Sez. di Padova, Padova, Italy}
\author{N.~Milas}
\affiliation{European Spallation Source, Box 176, SE-221 00 Lund, Sweden}
\author{R.~Moolya}
\affiliation{Institute for Experimental Physics, Hamburg University, 22761 Hamburg, Germany}
\author{J.L.~Mu\~noz}
\affiliation{Consorcio ESS-bilbao, Parque Cient\'{i}fico y Tecnol\'{o}gico de Bizkaia, Laida Bidea, Edificio 207-B, 48160 Derio, Bizkaia, Spain}
\author{K.~Niewczas}
\affiliation{Department of Physics and Astronomy, Ghent University, Proeftuinstraat 86, B-9000 Ghent, Belgium}
\author{M.~Oglakci}
\affiliation{University of Cukurova, Faculty of Science and Letters, Department of Physics, 01330 Adana, Turkey}
\author{T.~Ohlsson}
\email{Corresponding authors: S. Choubey (choubey@kth.se), T. Ohlsson (tohlsson@kth.se) and S. Vihonen (vihonen@kth.se)}
\affiliation{Department of Physics, School of Engineering Sciences, KTH Royal Institute of Technology, Roslagstullsbacken 21, 106 91 Stockholm, Sweden}
\affiliation{The Oskar Klein Centre, AlbaNova University Center, Roslagstullsbacken 21, 106 91 Stockholm, Sweden}
\author{M.~Olveg{\aa}rd}
\affiliation{Department of Physics and Astronomy, FREIA Division, Uppsala University, P.O. Box 516, 751 20 Uppsala, Sweden}
\author{M.~Pari}
\affiliation{Department of Physics and Astronomy "G. Galilei", University of Padova and INFN Sezione di Padova, Italy}
\author{D.~Patrzalek}
\affiliation{European Spallation Source, Box 176, SE-221 00 Lund, Sweden}
\author{G.~Petkov}
\affiliation{Sofia University St. Kliment Ohridski, Faculty of Physics, 1164 Sofia, Bulgaria}
\author{Ch.~Petridou}
\affiliation{Department of Physics, Aristotle University of Thessaloniki, Thessaloniki, Greece}
\affiliation{Center for Interdisciplinary Research and Innovation (CIRI-AUTH), Thessaloniki, Greece}
\author{P.~Poussot}
\affiliation{IPHC, Universit\'{e} de Strasbourg, CNRS/IN2P3, Strasbourg, France}
\author{A~Psallidas}
\affiliation{Institute of Nuclear and Particle Physics, NCSR Demokritos, Neapoleos 27, 15341 Agia Paraskevi, Greece}
\author{F.~Pupilli}
\affiliation{INFN Sez. di Padova, Padova, Italy}
\author{D.~Saiang}
\affiliation{Department of Civil, Environmental and Natural Resources Engineering Lule\aa~University~of~Technology, SE-971 87 Lulea, Sweden}
\author{D.~Sampsonidis}
\affiliation{Department of Physics, Aristotle University of Thessaloniki, Thessaloniki, Greece}
\affiliation{Center for Interdisciplinary Research and Innovation (CIRI-AUTH), Thessaloniki, Greece}
\author{A.~Scanu}
\affiliation{University of Milano-Bicocca and INFN Sez. di Milano-Bicocca, 20126 Milano, Italy}
\author{C.~Schwab}
\affiliation{IPHC, Universit\'{e} de Strasbourg, CNRS/IN2P3, Strasbourg, France}
\author{F.~Sordo}
\affiliation{Consorcio ESS-bilbao, Parque Cient\'{i}fico y Tecnol\'{o}gico de Bizkaia, Laida Bidea, Edificio 207-B, 48160 Derio, Bizkaia, Spain}
\author{G.~Stavropoulos}
\affiliation{Institute of Nuclear and Particle Physics, NCSR Demokritos, Neapoleos 27, 15341 Agia Paraskevi, Greece}
\author{R.~Tarkeshian}
\affiliation{European Spallation Source, Box 176, SE-221 00 Lund, Sweden}
\author{F.~Terranova}
\affiliation{University of Milano-Bicocca and INFN Sez. di Milano-Bicocca, 20126 Milano, Italy}
\author{T.~Tolba}
\affiliation{Institute for Experimental Physics, Hamburg University, 22761 Hamburg, Germany}
\author{M.~Topp-Mugglestone}
\affiliation{CERN, 1211 Geneva 23, Switzerland}
\author{E.~Trachanas}
\affiliation{European Spallation Source, Box 176, SE-221 00 Lund, Sweden}
\author{R.~Tsenov}
\affiliation{Sofia University St. Kliment Ohridski, Faculty of Physics, 1164 Sofia, Bulgaria}
\author{A.~Tsirigotis}
\affiliation{Physics Laboratory, School of Science and Technology, Hellenic Open University, 26335, Patras, Greece }
\author{S.~E.~Tzamarias}
\affiliation{Department of Physics, Aristotle University of Thessaloniki, Thessaloniki, Greece}
\affiliation{Center for Interdisciplinary Research and Innovation (CIRI-AUTH), Thessaloniki, Greece}
\author{M.~Vanderpoorten}
\affiliation{Department of Physics and Astronomy, Ghent University, Proeftuinstraat 86, B-9000 Ghent, Belgium}
\author{G.~Vankova-Kirilova}
\affiliation{Sofia University St. Kliment Ohridski, Faculty of Physics, 1164 Sofia, Bulgaria}
\author{N.~Vassilopoulos}
\affiliation{Institute of High Energy Physics (IHEP) Dongguan Campus, Chinese Academy of Sciences (CAS), Guangdong 523803, China}
\author{S.~Vihonen}
\email{Corresponding authors: S. Choubey (choubey@kth.se), T. Ohlsson (tohlsson@kth.se) and S. Vihonen (vihonen@kth.se)}
\affiliation{Department of Physics, School of Engineering Sciences, KTH Royal Institute of Technology, Roslagstullsbacken 21, 106 91 Stockholm, Sweden}
\affiliation{The Oskar Klein Centre, AlbaNova University Center, Roslagstullsbacken 21, 106 91 Stockholm, Sweden}
\author{J.~Wurtz}
\affiliation{IPHC, Universit\'{e} de Strasbourg, CNRS/IN2P3, Strasbourg, France}
\author{V.~Zeter}
\affiliation{IPHC, Universit\'{e} de Strasbourg, CNRS/IN2P3, Strasbourg, France}
\author{O.~Zormpa}
\affiliation{Institute of Nuclear and Particle Physics, NCSR Demokritos, Neapoleos 27, 15341 Agia Paraskevi, Greece}

\collaboration{ESSnuSB Collaboration}
\noaffiliation


\begin{abstract}
The ESSnuSB experiment aims to measure the leptonic CP phase $\delta_{CP}$ with an unprecedented resolution by probing neutrino oscillations at the second oscillation maximum. In the present work, the complementarity between the long-baseline neutrino program and atmospheric neutrinos is investigated for ESSnuSB. By simulating atmospheric neutrino events equivalent of 5.4 Mt$\cdot$year exposure, the resolution for $\delta_{\rm CP}^{}$ is found to improve from $7.5^\circ$ ($6.7^\circ$) to $7.1^\circ$ ($6.5^\circ$) at $1\sigma$~CL for $\delta_{\rm CP}^{} = -90^\circ$ ($+90^\circ$) with respect to super-beam neutrinos, resolving also the degeneracies arising from neutrino mass ordering. These findings highlight the synergies that exist between super-beam neutrinos and atmospheric neutrinos in ESSnuSB. 
\end{abstract}

\maketitle

\section{\label{sec:intro}Introduction}

It is well known that the present-day Universe is dominated by matter. In order for any theory to explain the asymmetry between baryons and antibaryons, it must meet the three Sakharov conditions~\cite{Sakharov:1967dj}. The European Spallation Source neutrino Super-Beam (ESSnuSB) experiment~\cite{ESSnuSB:2023ogw} in Sweden aims to investigate one of those conditions, {\em i.e.}, the violation of C and CP symmetries, by searching leptonic CP violation (CPV) near the second oscillation maximum. The experimental setup of ESSnuSB includes a neutrino superbeam facility at the European Spallation Source in Lund and a large neutrino detector complex that would be placed inside the mine in Zinkgruvan. The baseline length for such a setup would be 360~km. With 5~MW beam power and 540~kt fiducial mass, ESSnuSB could provide a sound experimental reach for the CP phase $\delta_{\rm CP}^{}$, which parametrizes CPV in the leptonic sector. The expected physics reach of ESSnuSB was first determined in the Conceptual Design Report~\cite{Alekou:2022emd}, where it is found that CPV could be established by at least $3\,\sigma$ confidence level (CL) for more than 70\% of the theoretically allowed values of $\delta_{\rm CP}^{}$. At the same time, $\delta_{CP}$ could be determined in ESSnuSB with a precision of $8^\circ$ ($1\,\sigma$~CL) or better, see also Refs.~\cite{ESSnuSB:2021azq,ESSnuSB:2024tmn}.

In addition to observing neutrinos from the accelerator source, the ESSnuSB far detectors (FD) would be suitable for detecting neutrinos from natural sources~\cite{Alekou:2022emd}. One example of such neutrinos are atmospheric neutrinos, which are created in cosmic-ray interactions inside the Earth's atmosphere. Most atmospheric neutrinos traverse long distances inside the Earth, which subjects atmospheric neutrinos to varying matter effects. For ESSnuSB, atmospheric neutrinos can be expected to provide complementary information to the neutrino beam program, especially on the neutrino mass ordering and the values of the neutrino oscillation parameters $\theta_{23}$ and $\Delta m_{31}^2$. It was found in Ref.~\cite{ESSnuSB:2024wet} that ESSnuSB would be able to determine the neutrino mass ordering by $3\sigma$~CL after 4 years of data taking and by $5\sigma$ after 10 years of data taking regardless of the true value of $\delta_{\rm CP}^{}$. At the same time, the precisions at which $\sin_{}^2\theta_{23}^{}$ and $\Delta m_{31}^2$ could be measured would improve notably from their presently known boundaries~\cite{Esteban:2024eli}. The prospects of observing atmospheric neutrinos at ESSnuSB FD were also studied for vector-mediated non-standard interactions~\cite{ESSnuSB:2025vsf}, for which the atmospheric neutrinos were shown to provide competitive sensitivities. In general, the information that could be acquired by observing atmospheric neutrinos at ESSnuSB FD is complementary to the physics prospects that can be expected for the super-beam,
which has a diverse physics program itself, {\em cf.}~Refs.~\cite{Blennow:2015nxa,Blennow:2019bvl,Ghosh:2019zvl,Choubey:2020dhw,Abele:2022iml,ESSnuSB:2023lbg,ESSnuSB:2024yji,Aguilar:2025lfm,ESSnuSB:2025shd} and~\cite{Majhi:2021api,Delgadillo:2023lyp,Capozzi:2023ltl,Cordero:2024nho}.

In the present work, we investigate potential synergies in measuring the value of $\delta_{\rm CP}^{}$ in the ESSnuSB physics program. This is achieved by generating Monte Carlo (MC) events for atmospheric neutrino interactions equivalent to $5.4$~Mt$\cdot$years. By co-analyzing neutrino oscillation data for both super-beam neutrinos and atmospheric neutrinos for the ESSnuSB setup, the sensitivity to $\delta_{\rm CP}^{}$ can be expected to improve. This is owed to the higher precision in measuring the values of $\sin_{}^2 \theta_{23}^{}$ and $\Delta m_{31}^2$ with atmospheric neutrinos for ESSnuSB FD in comparison to the long-baseline neutrino program in ESSnuSB, see {\em e.g.}~Ref.~\cite{Blennow:2019bvl}. In the present work, the synergies and sensitivities are examined thoroughly for $\delta_{\rm CP}^{}$ by simulating both atmospheric neutrinos and super-beam neutrinos for ESSnuSB. In this regard, the contribution from including atmospheric neutrinos in the numerical analysis are evaluated both on the CPV discovery potential and the $\delta_{\rm CP}^{}$ resolution. Moreover, the effects of angular resolution and energy resolution for atmospheric neutrinos are considered. It is found that the resolution on $\delta_{\rm CP}^{}$ is improved, while the CPV discovery potential remains mostly intact. On the other hand, the improved sensitivity to $\delta_{\rm CP}^{}$ leads to an improvement in the determination of neutrino mass ordering with the atmospheric neutrino sample~\cite{ESSnuSB:2024wet}. Therefore, the results obtained in this work highlight the complementarity from atmospheric neutrinos to the long-baseline neutrino program.

This article is divided into the five sections. In Section~\ref{sec:theory}, we discuss the theoretical aspects for measuring the neutrino oscillation parameters in long-baseline neutrino experiments for the standard three-flavor mixing. In that section, we revisit the importance of the first and second oscillation maximum. In Section~\ref{sec:methods}, the numerical methods are introduced for the simulation of super-beam neutrinos for the ESSnuSB setup. The simulation of super-beam neutrinos are carried out with \texttt{GLoBES}~\cite{Huber:2004ka,Huber:2007ji}. In the same section, we also review the numerical methods that are used in the analysis of atmospheric neutrino events, which in turn are generated within the \texttt{GENIE} framework~\cite{Andreopoulos:2009rq,GENIE:2021npt}. In Section~\ref{sec:results}, the main findings of this work are reported for analyzing the expected data for super-beam neutrinos and atmospheric neutrinos for the ESSnuSB setup. In Section~\ref{sec:concl}, we present the concluding remarks.

\section{\label{sec:theory}Theoretical overview}

Neutrino oscillation physics has witnessed nearly two decades of uninterrupted success in terms of experimental progress. The discovery of neutrino oscillations in early 2000s revealed that neutrinos have vanishingly small but non-zero mass. The mixing between neutrino masses and flavors is typically parametrized with the three mixing angles, $\theta_{12}^{}, \theta_{13}^{}$ and $\theta_{23}^{}$, the two mass-square differences, $\Delta m_{21}^2$ and $\Delta m_{21}^2$, and the leptonic CP phase $\delta_{\rm CP}^{}$. The mixing angles and the mass-square differences have all been measured to percent-level precision ($1\sigma$~CL) in the past and on-going neutrino oscillation experiments~\cite{Esteban:2024eli}, whereas the value of $\delta_{\rm CP}^{}$ is expected to be determined in future experiments, {\rm c.f.} Ref.~\cite{ESSnuSB:2023ogw}. The remaining unknowns in the three-flavor mixing scheme include the question on the ordering of neutrino masses, {\em i.e.}, whether the three neutrino masses $m_1^{}, m_2^{}$ and $m_3^{}$ follow normal ordering ($m_1^{} < m_2^{} < m_3^{}$, NO) or inverted ordering ($m_3^{} < m_1^{} < m_2^{}$, NO), and the uncertainty regarding the $\theta_{23}^{}$ octant. Moreover, it is not yet known whether the CP symmetry is violated in neutrino oscillations ($\sin \delta_{\rm CP}^{} \neq 0$) or if it is conserved ($\sin \delta_{\rm CP}^{} = 0$). Answers to those questions will be sought in the next-generation neutrino oscillation experiments, spearheaded by the long-baseline neutrino experiments T2HK~\cite{Hyper-KamiokandeProto-:2015xww} and DUNE~\cite{DUNE:2020ypp}, the reactor neutrino experiment JUNO~\cite{JUNO:2015zny}, and the neutrino telescopes Hyper-Kamiokande~\cite{Hyper-Kamiokande:2018ofw}, IceCube~\cite{IceCube-Gen2:2020qha}, INO~\cite{Thakore:2013xqa}, KM3NeT~\cite{KM3Net:2016zxf} and P-ONE~\cite{P-ONE:2020ljt}.

The experimental program of ESSnuSB employs a different strategy than what is planned for other long-baseline neutrino oscillation experiments, as the major goal for ESSnuSB is to study neutrino oscillations near the second oscillation maximum\footnote{Other proposed experiments that could grant access to the second oscillation maximum include MOMENT~\cite{Cao:2014bea} and T2HKK~\cite{Hyper-Kamiokande:2016srs}.}. The neutrino oscillations that could be studied in ESSnuSB can be described with the formula~\cite{Blennow:2019bvl} (see also Refs.~\cite{Freund:1999gy,Cervera:2000kp,Akhmedov:2004ny,Minakata:2015gra})
\begin{equation}
    \label{eq:probability}
        \begin{split}
        P_{\nu_\mu^{} \rightarrow \nu_e^{}}^{} &= s_{23}^2 \sin_{}^2 2\theta_{23}^{} \left( \frac{\Delta_{31}^{}}{B}\right)_{}^2 \sin_{}^2 \left( \frac{\tilde{B} L}{2} \right)\\
        & + c_{23}^2 \sin_{}^2 2\theta_{12}^{} \left( \frac{\Delta_{21}^{}}{A} \right)_{}^2 \sin_{}^2 \left( \frac{A L}{2} \right)\\
        & + \tilde{J} \frac{\Delta_{21}^{}}{A} \frac{\Delta_{31}^{}}{B} \sin \left( \frac{A L}{2} \right) \sin \left( \frac{\tilde{B} L}{2} \right)
        \left[ \cos \delta_{\rm CP}^{} \cos \left( \frac{\Delta_{31}^{} L}{2} \right) - \sin \delta_{\rm CP}^{} \sin \left( \frac{\Delta_{31}^{} L}{2} \right)\right],
        \end{split}
\end{equation}
where $s_{ij}^{} \equiv \sin \theta_{ij}^{}$ and $c_{ij}^{} \equiv \cos \theta_{ij}^{}$, $E_\nu^{}$ is the muon neutrino energy and $L$ is the baseline length. The matter effects are driven by $A \equiv \sqrt{2} G_{F}^{} N_{e}^{}$, where $G_{F}^{}$ and $N_{e}^{}$ are the Fermi coupling constant and the electron number density, respectively. Here, we have defined $\Delta_{ij}^{} = \Delta m_{ij}^2 / (2E_\nu^{})$, $\tilde{J} = c_{13}^{} \sin 2\theta_{12}^{} \sin 2\theta_{23}^{} \sin 2\theta_{13}^{}$ and $\tilde{B} \equiv |A + \Delta_{31}^{}|$. One can derive an analogous formula for $\bar{\nu}_\mu^{} \rightarrow \bar{\nu}_e^{}$ by replacing $\tilde{B}$ with $|A - \Delta_{31}^{}|$ on the first and third rows and inverting the sign of $\sin \delta_{\rm CP}^{} \sin \left( \Delta_{31}^{} L / 2\right)$ on the third row of equation~(\ref{eq:probability}).

The first term in equation~(\ref{eq:probability}) is driven by $\Delta_{31}^{}$ and is known as the ``atmospheric''  term~\cite{Blennow:2019bvl}. The second term in the equation is driven by $\Delta_{21}^{}$ and is denoted as the ``solar'' term. Finally, the third term in the equation depends both on $\Delta_{31}^{}$ and $\Delta_{21}^{}$. Therefore, we refer to the third as the ``interference'' term. For the ESSnuSB setup, the ``atmospheric'' term dominates the probability for $\nu_{\mu}^{} \rightarrow \nu_{e}^{}$ at the first oscillation maximum, whereas the ``interference'' term becomes the most significant near the second oscillation maximum~\cite{Blennow:2019bvl}. The ``solar'' term is mainly driven by matter effects and $\Delta_{21}^{}$, which means that the value of that term is strictly constrained by solar neutrino experiments. Measuring the value of $\delta_{\rm CP}^{}$ is therefore the most efficient near the second oscillation maximum, which is also targeted by the ESSnuSB proposal. In the following, we examine the differences between the first oscillation maximum and the second oscillation maximum more thoroughly.

\begin{figure}[!t]
    \centering
    \includegraphics[width=0.65\linewidth]{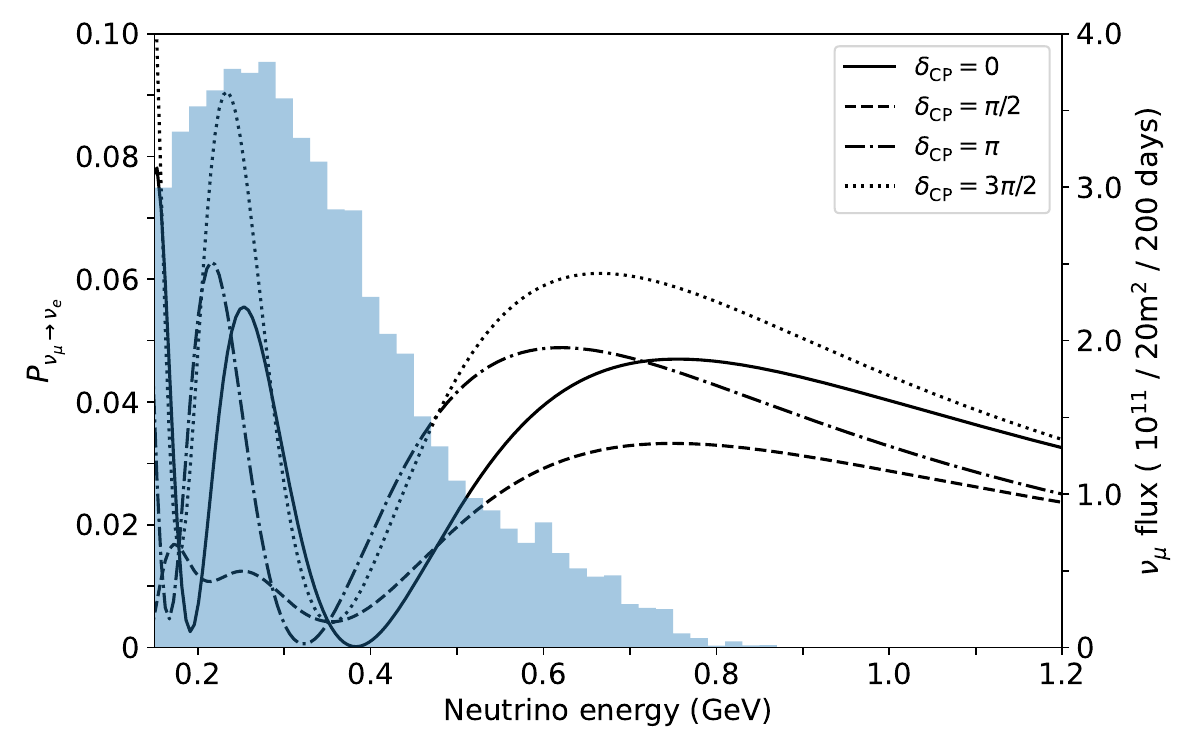}
    \caption{Neutrino oscillation probability $P_{\nu_{\mu}^{} \rightarrow \nu_{e}^{}}$ as a function of neutrino energy $E_\nu^{}$ for the ESSnuSB setup. The first and second oscillation maxima are found around $E_\nu^{} \sim 0.65$~GeV and $0.25$~GeV, respectively. The probability is shown for $\delta_{\rm CP}^{} = 0$ (solid curve)$,\delta_{\rm CP}^{} = \pi/2$ (dashed curve)$, \delta_{\rm CP}^{} = \pi$ (dot-dashed curve) and $\delta_{\rm CP}^{} = 3\pi/2$ (dotted curve). Neutrino mass ordering is assumed to be NO. The shaded histogram represents the $\nu_\mu^{}$ fluxes that have been obtained at $100$~km distance. The neutrino fluxes have been obtained from Ref.~\cite{Alekou:2022emd}.}
    \label{fig:probabilities}
\end{figure}

As ESSnuSB is expected to have a baseline length of $L = 360$~km and a neutrino beam spanning over neutrino energies $E_\nu^{} \in [0, 1.5]$~GeV, its configuration can be expected to be sensitive to the second oscillation maximum, and partially sensitive to the first oscillation maximum. The sensitivity to $\delta_{\rm CP}^{}$ is illustrated in Figure~\ref{fig:probabilities}, where the neutrino oscillation probability for the conversion $\nu_{\mu}^{} \rightarrow \nu_{e}^{}$ is presented as a function of neutrino energy $E_\nu^{}$ for the ESSnuSB setup. The average matter density for this case is $\rho \approx 2.7$~g/cm$^{-3}$. The probabilities are calculated with \texttt{GLoBES}. For this configuration, the first oscillation maximum occurs for the neutrino energies $E_\nu^{} \approx 0.65$~GeV, whereas the second oscillation maximum arises approximately at $E_\nu^{} \approx 0.25$~GeV. The probabilities for this neutrino oscillation channel are shown for $\delta_{\rm CP}^{} = 0$ (black solid curve), $\pi/2$ (black dashed curve), $\pi$ (black dot-dashed curve) and $3\pi/2$ (black dotted curve), which represent two possible CP-conserving values ($\delta_{\rm CP}^{} = 0$ or $\pi$) and two CP-violating values ($\delta_{\rm CP}^{} = \pi/2$ or $3\pi/2$). The $\nu_\mu^{}$ fluxes (shaded histogram) are also shown. The values of the neutrino oscillation parameters $\theta_{12}^{}, \theta_{13}^{}, \theta_{23}^{}, \Delta m_{21}^2$ and $\Delta m_{31}^2$ are adopted from the best-fit values of the recent global fit on experimental neutrino oscillation data~\cite{Esteban:2024eli,NuFIT:6.0}. For convenience, we have are listed the global best-fit values in Table~\ref{tab:bestfits}.

It is evident from Figure~\ref{fig:probabilities} that the first oscillation maximum spans over a wide range of neutrino energies, whereas the second oscillation maximum covers only a narrow window of neutrino energies. It can also be seen that varying the value of $\delta_{\rm CP}^{}$ causes larger fluctuations in the neutrino oscillation probability at the second oscillation maximum than at the first oscillation maximum. For the first oscillation maximum, where $E_\nu^{} \approx 0.65$~GeV, the difference between the probabilities obtained for $\delta_{\rm CP}^{} = 3\pi/2$ and $\delta_{\rm CP}^{} = \pi/2$ is $\Delta P_{\nu_\mu^{} \rightarrow \nu_e^{}}^{} \approx 0.03$. In contrast, near the second oscillation maximum, where $E_\nu^{} \approx 0.25$~GeV instead, the difference between the probabilities is much larger at $\Delta P_{\nu_\mu^{} \rightarrow \nu_e^{}}^{} \approx 0.08$. The stronger correlation between the probability and $\delta_{\rm CP}^{}$ near the second oscillation maximum arises from the interference term in equation~(\ref{eq:probability}).

\begin{figure}[!t]
    \centering
    \includegraphics[width=1.00\linewidth]{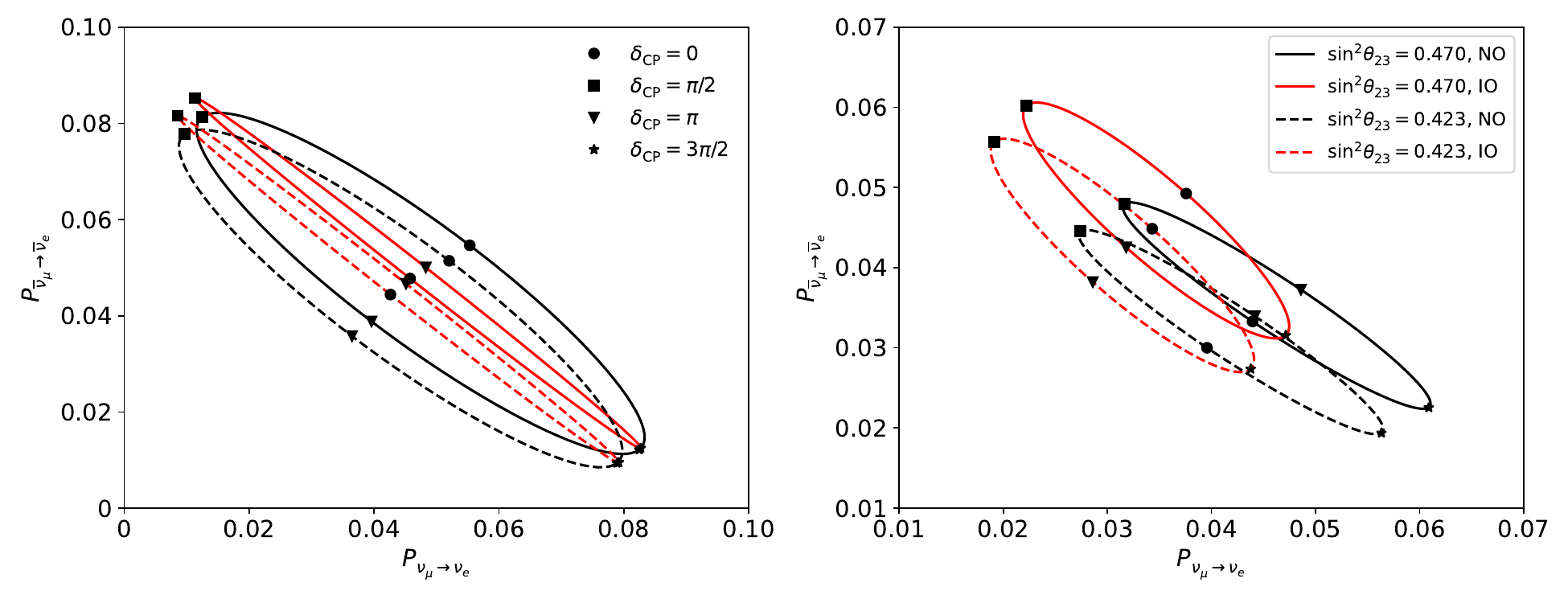}
    \caption{Correlation between the neutrino oscillation probabilities $P(\nu_\mu^{} \rightarrow \nu_e^{})$ and $P(\overline{\nu}_\mu^{} \rightarrow \overline{\nu}_e^{})$ for super-beam neutrinos with baseline length $L = 360$~km. The probabilities are obtained for neutrino energies $E_\nu = 0.25$~GeV (left panel) and $0.65$~GeV (right panel), which coincide with the second oscillation maximum and the first oscillation maximum, respectively. The correlations are shown for NO (black curves) and IO (red curves) at $\sin_{}^2 \theta_{23}^{} = 0.470$ (solid curves) and $\sin_{}^2 \theta_{23}^{} = 0.423$ (dashed curves). Note that the probabilities are shown up to $0.10$ ($0.07$) for the second (first) oscillation maximum.}
    \label{fig:biprobabilities}
\end{figure}

In order to understand the difference between neutrino oscillations in the first oscillation maximum and the second oscillation maximum, it is worth examining the correlation between the neutrino oscillation probability for $\nu_\mu^{} \rightarrow \nu_e^{}$ and $\bar\nu_\mu^{} \rightarrow \bar\nu_e^{}$ at the relevant neutrino energies. In Figure~\ref{fig:biprobabilities}, this correlation is presented by varying $\delta_{\rm CP}^{} \in [0, 2\pi)$ for the baseline length $L= 360$~km. The variation of $\delta_{\rm CP}^{}$ gives rise to ellipses due to the cyclic nature of $\delta_{\rm CP}^{}$. Depending on the neutrino energy at which the probabilities are obtained, the ellipses may coincide with an oscillation maximum or an oscillation minimum. In the left panel, the probabilities are displayed for the neutrino energy $E_\nu^{} = 0.25$~GeV, which matches the second oscillation maximum for $\nu_\mu^{} \rightarrow \nu_e^{}$. In contrast, the right panel shows the probabilities that are obtained at the first oscillation maximum, where $E_\nu^{} = 0.65$~GeV. The specific values $\delta_{\rm CP}^{} = 0, \pi/2, \pi$ and $3\pi/2$ are indicated by the circles, rectangles, triangles and stars, respectively. The probabilities are shown for four different cases: $\sin_{}^2\theta_{23}^{} = 0.470$ and mass ordering is NO (black solid curves), $\sin_{}^2\theta_{23}^{} = 0.723$ and mass ordering is NO (black dashed curves), $\sin_{}^2\theta_{23}^{} = 0.470$ and mass ordering is IO (red solid curves), and $\sin_{}^2\theta_{23}^{} = 0.423$ and mass ordering is IO (red dashed curves). Here $\sin_{}^2\theta_{23}^{} = 0.470$ matches with the central value for $\sin_{}^2 \theta_{23}^{}$ in Table~\ref{tab:bestfits}, whereas $\sin_{}^2\theta_{23}^{} = 0.423$ corresponds to $90\%$ of the same value.
\begin{table}[!t]
\begin{center}
\begin{tabular}{|c|c|c|c|}\hline
{\bf Parameter} & {\bf Central value} & {\bf Uncertainty (1\,$\sigma$~CL)} & {\bf Allowed range (3\,$\sigma$~CL)} \\ \hline
\rule{0pt}{3ex}$\sin^2 \theta_{12}^{}$ & $0.308$ & $3.8\%$ & $0.275 \rightarrow 0.345$ \\ 
\rule{0pt}{3ex}$\sin^2 \theta_{13}^{}$ & $0.02215$ & $2.7\%$ & $0.02030 \rightarrow 0.02388$ \\ 
\rule{0pt}{3ex}$\sin^2 \theta_{23}^{}$ & $0.470$ & $5.3\%$ & $0.435 \rightarrow 0.585$ \\ 
\rule{0pt}{3ex}$\delta_\text{CP}$ [$^\circ$] & $212$ & $18.9\%$ & $124 \rightarrow 364$ \\ 
\rule{0pt}{3ex}$\Delta m_{21}^2$ [10$^{-5}$ eV$^2$] & $7.49$ & $2.5\%$ & $6.92 \rightarrow 8.05$ \\ 
\rule{0pt}{3ex}$\Delta m_{31}^2$ [10$^{-3}$ eV$^2$] & $2.513$ & $0.8\%$ & $2.451 \rightarrow 2.578$ \\ \hline
\end{tabular}
\end{center}
\caption{\label{tab:bestfits} The best-fit values of the neutrino oscillation parameters as determined by neutrino experiments~\cite{Esteban:2024eli,NuFIT:6.0}. The values and their relative uncertainties are presented at $1\sigma$~CL assuming normal mass ordering. The allowed values are also shown for each parameter at $3\sigma$~CL. The values are based on \texttt{NuFit~6.0} best-fit results~\cite{NuFIT:6.0} including Super-Kamiokande data.}
\end{table}

In Figure~\ref{fig:biprobabilities}, the phenomenological difference in the physics reach near the first oscillation maximum and the second oscillation maximum is illustrated. At the second oscillation maximum (left panel), the ellipses that correspond to NO and IO are mostly overlapping. Determining the value of $\delta_{\rm CP}^{}$ is easier for this case, since the ellipses extend over a wide area and $\delta_{\rm CP}^{}$ values become easier to distinguish. However, identifying the correct ordering of neutrino masses at the same time is challenging, as there is no clear separation between the black and red ellipses. If the value of $\sin_{}^2\theta_{23}^{}$ were allowed to vary, the red ellipses would become entirely covered by the black ellipses, which would consequently make the determination of $\delta_{\rm CP}^{}$ impossible. This issue is not present for the first oscillation maximum (right panel), where the ellipses are clearly separated. In contrast, the probabilities span over smaller ranges, which means that determining the value of $\delta_{\rm CP}^{}$ becomes more challenging in comparison to the second oscillation maximum. The separation between the black ellipses and the red ellipses at the first oscillation maximum arises from matter effects, which bears resemblance to CPV. For the ESSnuSB setup, the matter effects have relatively small impact due to the comparatively short baseline length, and the separation of the ellipses corresponding to the NO case and the IO case is not significant. This is also apparent in the right panel of Figure~\ref{fig:biprobabilities}, where the black and red ellipses overlap by about $50\%$ of their respective surface areas. Therefore, the experimental setup of ESSnuSB would not have any significant advantage to determine neutrino mass ordering with its super-beam neutrinos.

While the prospects to determine the neutrino mass ordering at ESSnuSB are rather small even for the first oscillation maximum, the sensitivity to the value of $\delta_{\rm CP}^{}$ is very promising. This can be confirmed with the left panel of Figure~\ref{fig:biprobabilities}, where the ellipses corresponding to the different values of $\sin_{}^{2} \theta_{23}^{}$ are separated by a linear shift. For both NO and IO, the CP-conserving values $\delta_{\rm CP}^{} = 0$ and $\delta_{\rm CP}^{} = \pi$ are located along a straight line that can be drawn through the circles and triangles. The same is true for the CP-violating value $\delta_{\rm CP}^{} = \pi/2$ and also for $\delta_{\rm CP}^{} = 3\pi/2$, which are located at the opposite sides of the ellipses. If the value of $\sin_{}^2 \theta_{23}^{}$ is varied over an interval, the location of a certain value of $\delta_{\rm CP}^{}$ could always be determined at the same part of the ellipses. This also indicates that the second oscillation maximum is able to provide a higher sensitivity to the value of $\delta_{\rm CP}^{}$ than the first oscillation maximum, where the ellipses span over a narrower range of probabilities and the different values of $\delta_{\rm CP}^{}$ are located closer to each other. It is important to notice that the ellipses shown in the left panel of Figure~\ref{fig:biprobabilities} and much larger than the corresponding ellipses shown in the right panel. This observation indicates that the resolution for $\delta_{\rm CP}^{}$ can be expected to be substantially better for the second oscillation maximum. As the proposed ESSnuSB setup has a better reach for neutrino oscillations at the second oscillation maximum, {\em cf.}~Figure~\ref{fig:probabilities}, ESSnuSB can also be expected to measure the value of $\delta_{\rm CP}^{}$ with a high precision.

Atmospheric neutrinos can be expected to contribute to the $\delta_{\rm CP}^{}$ measurement by constraining the experimental limits on $\sin_{}^2 \theta_{23}^{}$ and $\Delta m_{31}^2$. Thanks to their exposure to very strong matter effects, atmospheric neutrinos are also sensitive to neutrino mass ordering and $\theta_{23}^{}$ octant, see Ref.~\cite{ESSnuSB:2024wet}. In the remainder of this article, we discuss the prospects of observing atmospheric neutrinos at ESSnuSB FD and study their complementarity to the long-baseline neutrino program at ESSnuSB. 

\section{\label{sec:methods}Simulation methods}

The simulation of super-beam neutrinos for the ESSnuSB setup is carried out entirely within the \texttt{GLoBES} framework~\cite{Huber:2004ka,Huber:2007ji}. The analysis of the super-beam neutrinos is done with the following $\chi^2_{}$ function:
\begin{equation}
    \label{eq:chi2_beam}
        \chi_{\rm beam}^2 = 2 \sum_{i=1}^{n} \left( N_i^{\rm test} - N_i^{\rm true} + N_i^{\rm true} \log \frac{N_i^{\rm true}}{N_i^{\rm test}} \right) + \frac{\zeta_{\rm sg}^2}{\sigma_{\rm sg}^2} + \frac{\zeta_{\rm bg}^2}{\sigma_{\rm bg}^2},
\end{equation}
where the index $i = 1, \ldots, n$ runs over the analysis bins that contain the super-beam neutrino events, while $n$ denotes the number of bins for each channel. The super-beam neutrino events stem from the neutrino oscillation channels $\nu_{\mu}^{} \rightarrow \nu_{e}^{}, \nu_{\mu}^{} \rightarrow \nu_{\mu}^{}, \bar{\nu}_{\mu}^{} \rightarrow \bar{\nu}_{e}^{}$ and $\bar{\nu}_{\mu}^{} \rightarrow \bar{\nu}_{e}^{}$. Here $N_{i}^{\rm true}$ and $N_{i}^{\rm test}$ are the neutrino events corresponding to the true and test values of the neutrino oscillation parameters, respectively. Neutrino events are computed as a sum of the corresponding signal and background events. The systematic uncertainties pertaining to the signal and background events are parameterized with the pull-parameters $\zeta_{\rm sg}^{}$ and $\zeta_{\rm bg}^{}$, which describe uncorrelated normalization errors with standard deviations $\sigma_{\rm sg}^{}$ and $\sigma_{\rm bg}^{}$. See Ref.~\cite{Fogli:2002pt} for more details on the implementation of the pull-method.

The calculation of the neutrino events $N_i^{\rm test}$ and $N_i^{\rm true}$ is carried out in \texttt{GLoBES} using the same experiment simulation setup that was adopted for the ESSnuSB Conceptual Design Report~\cite{Alekou:2022emd}. In that setup, the systematic uncertainties are modeled with $5\%$ normalization errors for the signal events and $5\%$ normalization errors for the background events, respectively. In addition to the normalization error, the simulation of the ESSnuSB setup includes systematic uncertainties for the energy calibration of its neutrino detectors and also an uncorrelated bin-to-bin error, which are taken into account in migration matrices. The expected running time for the neutrino beam is 10 years, which would be divided into 5 years of running in the muon neutrino production mode and 5 years in the muon antineutrino production mode. Each of the studied neutrino oscillation channels is given its own nuisance parameters to account the systematic uncertainties for the signal and background events. The event rate calculation utilizes the same neutrino fluxes, cross-sections and migration matrices that were used to produce results in Ref.~\cite{Alekou:2022emd}.

The simulation of atmospheric neutrinos for ESSnuSB FD is based on Ref.~\cite{ESSnuSB:2024wet}. The generation of atmospheric neutrino events for the ESSnuSB far detector is done with the corresponding MC event generator in \texttt{GENIE}~\cite{Andreopoulos:2009rq,GENIE:2021npt}, which utilizes \texttt{ROOT}~\cite{Brun:1997pa}. The detector geometry for ESSnuSB FD is defined as a preliminary \texttt{ROOT} geometry that corresponds to two cylindrical vessels of ultra-pure water with 540~kt total fiducial mass. The atmospheric neutrino fluxes are adopted from the Honda simulation~\cite{Honda:2015fha} for Pyh\"asalmi, for which the fluxes have been averaged over the full solar cycle. To minimize the effect of MC fluctuations in the analysis, atmospheric neutrino interactions are simulated for an exposure that corresponds to 500 years of data taking. The MC events are generated for neutrino energies $E_\nu^{} \in [0.1, 100]$~GeV. The generated MC events are normalized for 10 years, which is equivalent to 5.4~Mt$\cdot$year exposure. The analysis of the MC events is done with the $\chi^2_{}$ function
\begin{equation}
    \label{eq:chi2_atmospheric}
    \chi_{\rm atmospheric}^2 = 2 \sum_{j=1}^{2000} \left( E_j^{} - O_j^{} + O_j^{} \log \frac{O_j^{}}{E_j^{}} \right) + \sum_{k=1}^{5} \left( \frac{\zeta_k^{}}{\sigma_k^{}} \right)^2,
\end{equation}
where $j = 1,\ldots,2000$ is the number of the analysis bin for atmospheric neutrinos and $E_j^{}$ is the number of expected events that are expected for the test hypothesis. Correspondingly, $O_j^{}$ provides the number of observed events, which are computed for the truth hypothesis. The analysis bins contain the information from the electron-like events and muon-like events, which comprise of events that involve $\nu_e^{}$ and $\bar\nu_e^{}$ interactions and $\nu_\mu^{}$ and $\bar\nu_\mu^{}$ interactions, respectively. The sampled events are multiplied with appropriate detector efficiencies, which are adopted from Ref.~\cite{Alekou:2022emd}. The average values for the detector efficiencies are $62\%$, $54\%$, $69\%$ and $68\%$ for $\nu_\mu^{}, \nu_e^{}, \bar\nu_\mu^{}$ and $\bar\nu_{e}^{}$ events, respectively. Following the example of our previous studies~\cite{ESSnuSB:2024wet,ESSnuSB:2025vsf}, the energy and angular resolution for atmospheric neutrino events are implemented by applying Gaussian smearing for neutrino energies and zenith angles. In this regard, a $30\%$ ($10\%$) energy resolution is assumed for MC events falling into sub-GeV (multi-GeV) bins, while $10^\circ$ angular resolution is assumed for all MC events.

Atmospheric neutrinos are sensitive to the neutrino oscillation channels $\nu_{\mu}^{} \rightarrow \nu_\mu^{}, \nu_{e}^{} \rightarrow \nu_{\mu}^{}$ and $\nu_{\mu}^{} \rightarrow \nu_{e}^{}$ and their antineutrino counterparts, but without the sensitivity to electric charge of the charged leptons\footnote{There is an ongoing study on implementing gadolinium-doping in ESSnuSB FD.}. The oscillation effects are taken into account in the event samples with a reweighting mechanism~\cite{ESSnuSB:2024wet}, which also obtains its neutrino oscillation probabilities from \texttt{GLoBES}. In this regard, the matter density profile is obtained using the Preliminary Reference Earth Model (PREM)~\cite{Dziewonski:1981xy} with 81 layers. For each MC event, a random number $S \in [0, 1]$ is generated and compared with relevant neutrino oscillation probabilities. The result determines whether the atmospheric neutrino in question has oscillated to another flavor state or stayed in the same flavor state.

The systematic uncertainties for the atmospheric neutrino events~\cite{Gonzalez-Garcia:2004pka,Gandhi:2007td,Ghosh:2012px,Choubey:2015xha} are described with the pull-parameters $\zeta_{k}^{} = \zeta_1^{}, \ldots, \zeta_5^{}$, which model the systematic uncertainties related to the atmospheric neutrinos, while $\sigma_{k}^{}$ indicates their relative uncertainties. The systematic uncertainties alter the expected events $E_j^{}$ as
\begin{equation}
    \label{eq:atmospheric_events}
    E_j^{} = E_{j, 0}^{} \left( 1 + \sum_{k=1}^{5} f_{j, k}^{} \zeta_k^{} \right),
\end{equation}
where $E_{j, 0}^{}$ is the unaltered number of atmospheric neutrino events.

There are five types of systematic uncertainties that are taken into account for atmospheric neutrino simulations. Those systematic uncertainties are summarized in Table~\ref{tab:systematics}, where each systematic uncertainty is assigned one pull-parameter. Electron-like events and muon-like events are treated with separate sets of nuisance parameters. Following the example of our previous studies~\cite{ESSnuSB:2024wet,ESSnuSB:2025vsf}, we adopt $20\%$ uncertainty for the atmospheric neutrino fluxes, $10\%$ uncertainty for the atmospheric neutrino cross-sections, and $5\%$ uncertainty for the neutrino detector efficiencies. The uncertainty related to the neutrino cosine zenith angle is computed as $5\%$ of the average neutrino cosine zenith angle $\langle\cos \theta_{z}^{}\rangle$ in the analysis bin. In contrast, the systematic uncertainty describing the potential deviations from the power-law spectrum in atmospheric neutrino fluxes is computed by introducing a $5\%$ variation to the power-law spectrum that is used to describe atmospheric neutrino fluxes. The tilted power-law spectrum is obtained from the unaltered atmospheric neutrino fluxes as~\cite{Gonzalez-Garcia:2004pka}
\begin{equation}
    \label{eq:tilt_spectrum}
        \Phi_{\delta}^{} (E_{\nu}^{}) = \Phi_{\nu, 0}^{} \left(\frac{E_{\nu}}{E_{0}}\right)_{}^{\delta} \simeq \Phi_{\nu, 0}^{} \left( 1 + \delta \log \frac{E_{\nu}^{}}{E_{0}^{}} \right),
\end{equation}
where $\Phi_{\nu, 0}^{}$ is the original atmospheric neutrino flux. The perturbation to the atmospheric neutrino fluxes is introduced with the parameter $\delta$ and reference energy $E_{0}^{}$. The weights for the systematic uncertainties responsible for the tilt error are computed as the fraction of the MC events that are generated for the tilted and unaltered atmospheric neutrino fluxes. For the present work, the atmospheric neutrino fluxes are adopted from Ref.~\cite{Honda:2015fha} for Pyh\"asalmi and the energy tilt error is obtained with the perturbation parameter $\delta = 5\%$ and the reference energy $E_{0}^{} = 2~{\rm GeV}$. As Pyh\"asalmi is located relatively close to Zinkgruvan, the fluxes from Ref.~\cite{Honda:2015fha} are also sufficient to approximate the conditions for ESSnuSB FD.
\begin{table}[!t]
\begin{center}
\begin{tabular}{|c|c|}\hline
{\bf Systematic error} & {\bf Uncertainty} \\ \hline
\rule{0pt}{3ex}Flux normalization & $20\%$ \\
\rule{0pt}{3ex}Cross-section normalization & $10\%$ \\
\rule{0pt}{3ex}Zenith angle dependence & varies \\
\rule{0pt}{3ex}Energy tilt & varies \\
\rule{0pt}{3ex}Detector & $5\%$ \\ \hline
\end{tabular}
\end{center}
\caption{\label{tab:systematics}List of systematic uncertainties used in this work. See the text for the implementation of the zenith angle dependence and energy tilt errors. More information on the methodology can be found in Ref.~\cite{Gonzalez-Garcia:2004pka,Gandhi:2007td,Ghosh:2012px,Choubey:2015xha}.}
\end{table}

The main sources of background for atmospheric neutrinos are constituted by atmospheric muons and beam-rock muons~\cite{Alekou:2022emd,ESSnuSB:2024wet}. As the mine in Zinkgruvan is approximately 1000~m deep, the atmospheric muon background is significantly reduced by the time they reach ESSnuSB FD~\cite{Alekou:2022emd}. On the other hand, the relatively low energy of beam neutrinos in ESSnuSB ensures that the majority of beam-rock muons created in ESSnuSB fall below the energy threshold set for atmospheric neutrino detection at ESSnuSB FD. Therefore, the effects of the atmospheric muon and beam-rock muon backgrounds can be safely omitted in the atmospheric neutrino analysis. The background events that are relevant for the neutrino super-beam comprise purely of beam-related backgrounds, whereas external backgrounds can be rejected with the beam-time window method~\cite{Alekou:2022emd}.

The joint analysis of the super-beam and atmospheric neutrino data for ESSnuSB is carried out as follows. The $\chi_{}^2$ function adopted for the analysis is given by
\begin{equation}
    \label{eq:chi2_beam_and_atmospheric}
        \chi_{}^2 = \chi_{\rm beam}^2 + \chi^2_{\rm atmospheric} + \chi^2_{\rm priors},
\end{equation}
where $\chi_{\rm beam}^2$ and $\chi_{\rm atmospheric}^2$ are calculated according to equations~(\ref{eq:chi2_beam}) and (\ref{eq:chi2_atmospheric}), respectively. External priors for the neutrino oscillation parameters are included in $\chi_{\rm priors}^2$. In the numerical analysis, the $\chi_{}^2$ function defined in equation~(\ref{eq:chi2_beam_and_atmospheric}) is minimized over the oscillation parameters $\theta_{12}^{}, \theta_{13}^{}, \theta_{23}^{}, \delta_{\rm CP} ^{}, \Delta m_{21}^2$ and $\Delta m_{31}^2$. The minimization is carried out either with or without knowledge on neutrino mass ordering. Since atmospheric neutrino oscillations are mainly sensitive to variations on $\theta_{23}^{}$ and $\Delta m_{31}^2$, and to a smaller extent on $\delta_{\rm CP}^{}$, the atmospheric neutrino contribution $\chi_{\rm atmospheric}^2$ is provided as a function of those parameters. For the combined analysis of the super-beam and atmospheric neutrino events, we adopt Gaussian priors for $\theta_{12}^{}, \theta_{13}^{}$ and $ \Delta m_{21}^2$. Whenever super-beam neutrinos are analysed without atmospheric neutrinos, Gaussian priors are included for $\theta_{23}^{}$ and $\Delta m_{31}^2$ as well. In both cases, central values and standard deviations for the priors are adopted from Table~\ref{tab:bestfits}. For concreteness, all results presented in this article are obtained assuming the true neutrino mass ordering to be NO, as suggested by the latest global fit on experimental neutrino data~\cite{NuFIT:6.0}.

\section{\label{sec:results}Numerical results}

\begin{figure}[!t]
    \centering
    \includegraphics[width=1.00\linewidth]{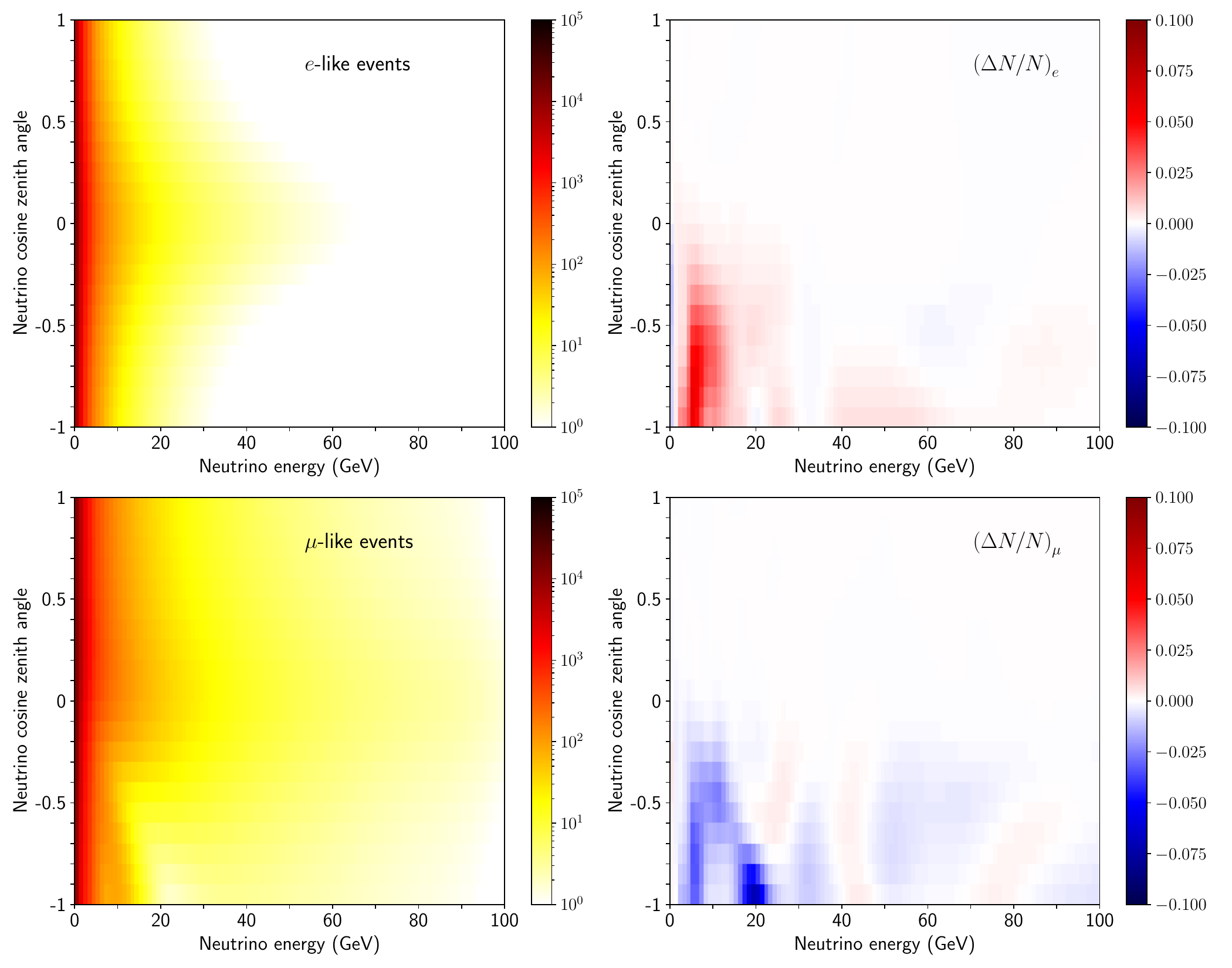}
    \caption{Atmospheric neutrino events expected for ESSnuSB FD. The expected $e$-like event (upper left) and $\mu$-like event (lower left) samples are shown for an equivalent of 5.4~Mt$\cdot$year exposure taking into account the detector effects. The relative differences in events computed for $\sin_{}^2 \theta_{23}^{} = 0.470$ and $\sin_{}^2 \theta_{23}^{} = 0.517$ are displayed for $e$-like events (upper right) and $\mu$-like events (lower right), where $\sin_{}^2 \theta_{23}^{} = 0.470$ is the central value. Neutrino mass ordering is assumed to be NO.}
    \label{fig:events}
\end{figure}

Complementarity between atmospheric neutrinos and super-beam neutrinos can be illustrated by comparing neutrino events. In Figure~\ref{fig:events}, the expected number of atmospheric neutrino events is presented as a function of neutrino energy bins and neutrino cosine zenith angle bins for electron-like events (top-left panel) and muon-like events (bottom-left panel). Here, electron-like events are composed of the generated atmospheric neutrino events where interaction with either $\nu_e^{}$ or $\bar\nu_e^{}$ is involved. Similarly, muon-like events consist of atmospheric neutrino events where either $\nu_\mu^{}$ or $\bar\nu_\mu^{}$ takes part in the interaction. The detector effects are taken into account as described in Section~\ref{sec:methods}. Moreover, neutrino oscillations are taken into account in the electron-like events and muon-like events assuming the global best-fit values for the neutrino oscillation parameters, see Table~\ref{tab:bestfits} in Section~\ref{sec:theory}. The dependence on the value of $\sin_{}^2 \theta_{23}^{}$ is shown for the electron-like (top-right panel) and muon-like samples (bottom-right panel), where the atmospheric neutrino events are simulated for the central value ($\sin_{}^2 \theta_{23}^{} = 0.470$) and a slightly elevated value ($\sin_{}^2 \theta_{23}^{} = 0.517$). The resulting changes in the expected electron-like and muon-like samples are shown as $(\Delta N / N)_e \equiv (N_e^{\rm higher} - N_e^{\rm center}) / N_e^{\rm center}$ and $(\Delta N / N)_\mu \equiv (N_\mu^{\rm higher} - N_\mu^{\rm center}) / N_\mu^{\rm center}$, where the superscripts {\em higher} and {\em center} indicate that the neutrino oscillation probabilities are calculated for $\sin_{}^2 \theta_{23}^{} = 0.517$ and $\sin_{}^2 \theta_{23}^{} = 0.470$, respectively. Note that $\sin_{}^2 \theta_{23}^{} = 0.517$ is $10\%$ higher than the central value $\sin_{}^2 \theta_{23}^{} = 0.470$. The computed event differences in Figure~\ref{fig:events} show that electron-like events appear for a number of neutrino energy bins, especially around neutrino energies $E_\nu^{} \sim 6$~GeV as a result of $\nu_\mu^{} \rightarrow \nu_e^{}$ oscillations. At the same time, muon-like events disappear due to neutrino oscillations in the $\nu_\mu^{} \rightarrow \nu_e^{}$ and $\nu_\mu^{} \rightarrow \nu_\tau^{}$ channels. Neutrino oscillations through $\nu_\mu^{} \rightarrow \nu_\tau^{}$ are particularly significant around neutrino energies $E_\nu^{} \sim 20$~GeV. Smaller changes in the electron-like and muon-like events can be observed higher neutrino energies. The neutrino oscillation probabilities are computed assuming neutrino mass ordering to be NO, which is currently favored by the global neutrino oscillation data. It is indicated by Figure~\ref{fig:events} that atmospheric neutrinos are very sensitive to the value of $\sin_{}^2 \theta_{23}^{}$. One can similarly obtain relative event differences for a case where the value of $\Delta m_{31}^2$ is varied instead of $\sin_{}^2 \theta_{23}^{}$. These analyses at event level indicate that atmospheric neutrinos have a strong dependence on the values of $\sin_{}^2 \theta_{23}^{}$ and $\Delta m_{31}^2$.

\begin{figure}[!t]
    \centering
    \includegraphics[width=1.00\linewidth]{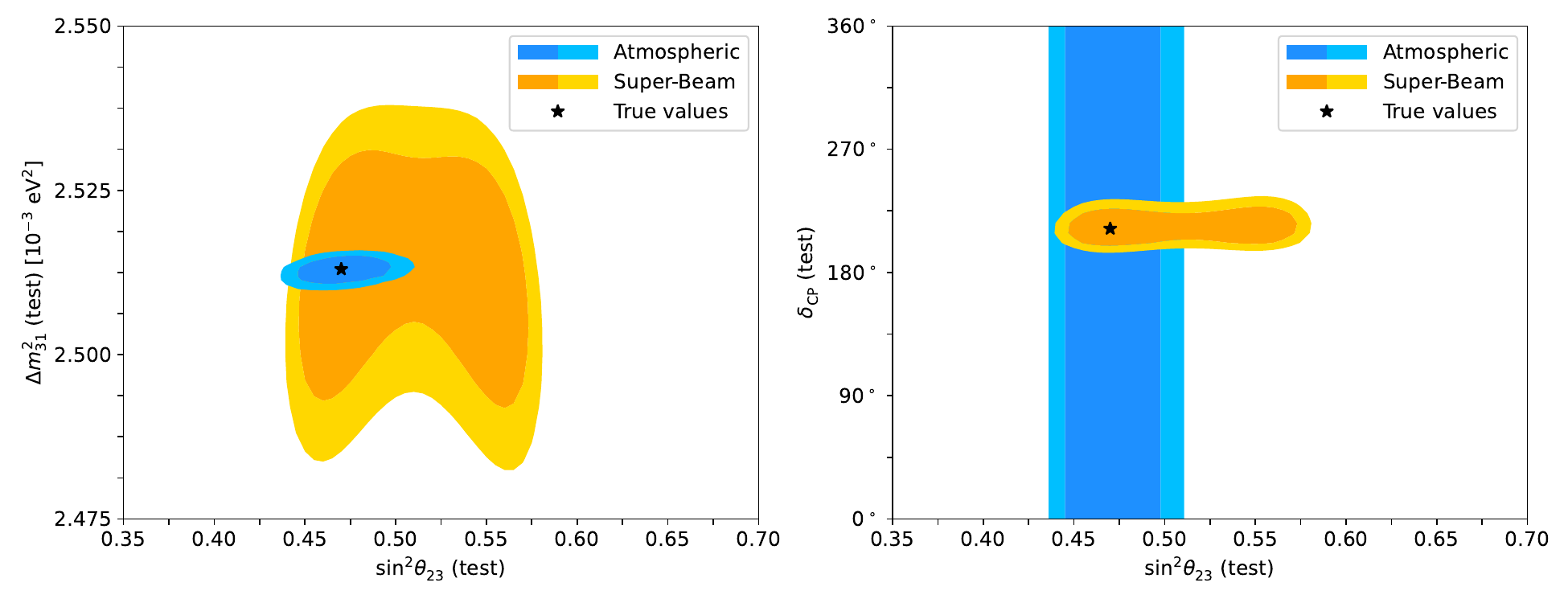}
    \caption{Allowed values for $\sin_{}^{2}\theta_{23}^{}$ and $\Delta m_{31}^2$ (left panel) and for $\sin_{}^{2}\theta_{23}^{}$ and $\delta_{\rm CP}^{}$ (right panel) by analyzing data for atmospheric neutrinos (blue regions) and super-beam neutrinos (yellow regions) at ESSnuSB FD (2 dof). The allowed values are shown at $2\sigma$~CL (darker colors) and $3\sigma$~CL (lighter colors), while the true values for $\sin_{}^{2} \theta_{23}^{}$ and $\Delta m_{31}^2$ are indicated by the black stars. The true neutrino mass ordering is assumed to be NO. Note that no priors are used for $\sin_{}^2 \theta_{23}^{}, \Delta m_{31}^2$ or $\delta_{\rm CP}^{}$.}
    \label{fig:allowed_values}
\end{figure}

Atmospheric neutrinos can provide complementary information to the neutrino super-beam by improving the precisions on $\sin_{}^2 \theta_{23}^{}$ and $\Delta m_{31}^2$. The sensitivity to these two quantities is shown in Figure~\ref{fig:allowed_values}. The figure shows the allowed parameter space for $\sin_{}^2 \theta_{23}^{}$ and $\Delta m_{31}^2$ (left panel) after ESSnuSB far detector has been collecting data for atmospheric neutrinos for 10 years (blue region). The values that are allowed for $\sin_{}^{2}\theta_{23}^{}$ and $\Delta m_{31}^2$ by $3\sigma$~CL are indicated with the blue region. Atmospheric neutrinos are sensitive to the values of $\sin_{}^2 \theta_{23}^{}$ and $\Delta m_{31}^2$ through the access to the first oscillation maximum. In contrast, super-beam neutrinos (yellow region) with the same running time are not as sensitive to the first oscillation maximum. Therefore, the sensitivity that can be gained from the atmospheric neutrino sample for those two observables is higher in comparison to the super-beam neutrino sample. This is evident from Figure~\ref{fig:allowed_values}, where the allowed region for super-beam neutrinos covers a larger area than the values that would be allowed by the atmospheric neutrino sample. The results have been obtained by assuming the neutrino mass ordering to be NO. Moreover, the corresponding fit results are shown for $\sin_{}^2 \theta_{23}^{}$ and $\delta_{\rm CP}^{}$ (right panel). For $\delta_{\rm CP}^{}$, atmospheric neutrinos have no apparent sensitivity, as the allowed value region seems to be independent of the value of $\delta_{\rm CP}^{}$. This behavior can also be traced to the dominance of the first oscillation maximum for atmospheric neutrinos, for which the value of $\delta_{\rm CP}^{}$ has an insignificant effect. In comparison, the allowed value region that can be obtained for super-beam neutrinos sets stringent constraints on $\delta_{\rm CP}^{}$, despite their lower sensitivity to $\sin_{}^2 \theta_{23}^{}$.

The synergy that exists between super-beam neutrinos and atmospheric neutrinos for the ESSnuSB setup is evident. In Figure~\ref{fig:allowed_values}, the comparatively stronger dependence on $\sin_{}^{2} \theta_{23}^{}$ and $\Delta m_{31}^{2}$ for the atmospheric neutrino sample indicates that the sensitivity to the first oscillation maximum can provide insight that is not easily accessible for the super-beam neutrino sample. The most striking result is related to the sensitivity to $\Delta m_{31}^2$, for which atmospheric neutrinos are able to provide noticeably higher precision than super-beam neutrinos. This can be seen by comparing the blue and yellow color regions in the left panel of Figure~\ref{fig:allowed_values}. The strong connection to the value of $\Delta m_{31}^2$ arises from the access to the first oscillation maximum, which occurs at multi-GeV energies for atmospheric neutrinos. For the same reason, the atmospheric neutrino sample would be more efficient in constraining the allowed values for $\sin_{}^2 \theta_{23}^{}$, while the sensitivity to $\delta_{\rm CP}^{}$ appears negligible for atmospheric neutrinos.

\begin{figure}[!t]
    \centering
    \includegraphics[width=1.0\linewidth]{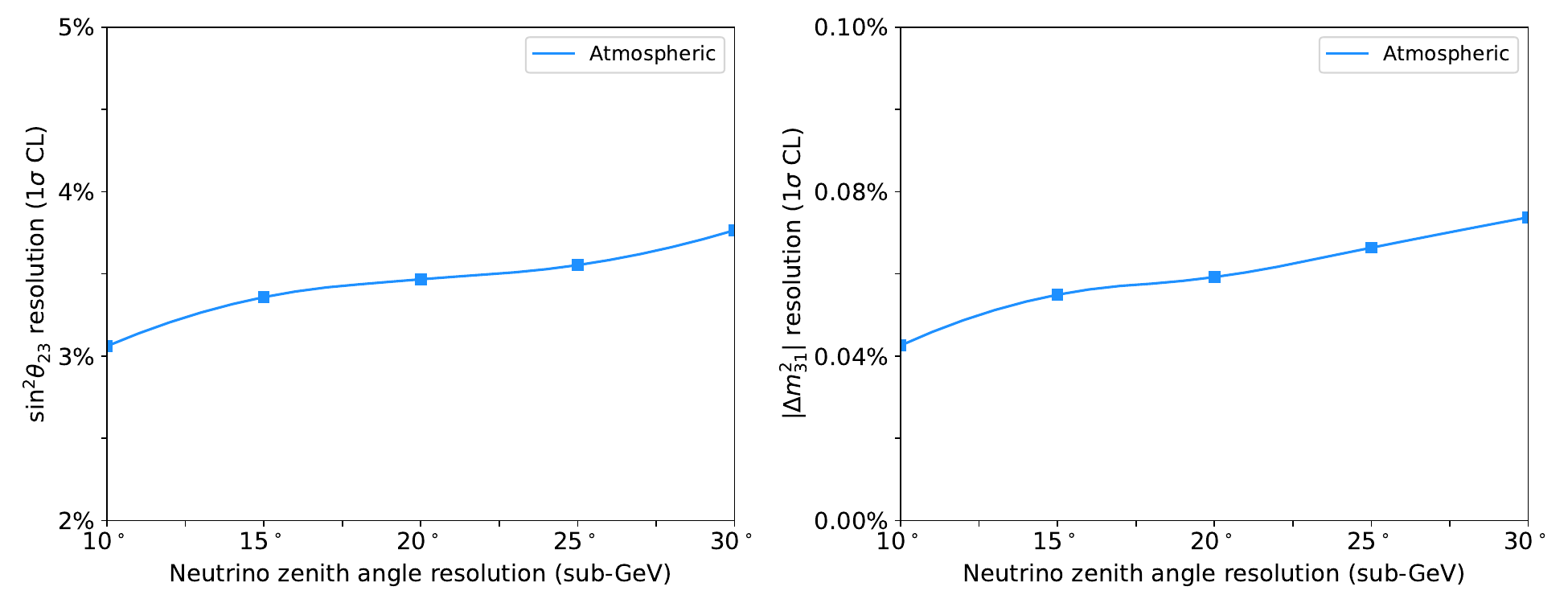}
    \caption{Effect of neutrino zenith angle resolution on the precisions at which $\sin^2_{} \theta_{23}^{}$ (left panel) and $|\Delta m_{31}^2|$ (right panel) could be measured by analyzing atmospheric neutrino data for ESSnuSB FD (2 dof). The neutrino zenith angle resolution is varied for the MC events in the sub-GeV neutrino energy bins. The results are indicated by the blue rectangles, which have been fitted with cubic splines. The true neutrino mass ordering is assumed to be NO.}
    \label{fig:resolutions}
\end{figure}

The sensitivity to neutrino oscillation parameters by analyzing the atmospheric neutrino sample depends largely on the neutrino detector's capabilities to reconstruct the neutrino energy and the neutrino zenith angle for each event. While neutrino energy and neutrino zenith angle have relatively strong correlations with those of their respective outgoing charged lepton for multi-GeV neutrino energies, the energy and zenith angle of atmospheric neutrinos becomes more uncertain for events with sub-GeV neutrino energies. In Figure~\ref{fig:resolutions}, the effect of the neutrino zenith angle resolution in the sub-GeV neutrino energy bins is quantified for $\sin_{}^2 \theta_{23}^{}$ and $\Delta m_{31}^2$. 

In Figure~\ref{fig:resolutions}, the expected resolutions for $\sin_{}^{2} \theta_{23}^{}$ (left panel) and $\Delta m_{31}^{2}$ at $1\sigma$~CL (right panel) are shown as functions of the neutrino zenith angle resolution that is assumed for the reconstruction of the atmospheric neutrino events in the sub-GeV neutrino energy bins. The effect of the neutrino zenith angle resolution is calculated assuming the true mass ordering to be NO. The blue rectangles represent the expected $1\sigma$~CL uncertainties for $\sigma(\cos\theta_{z}^{}) = 10^\circ, 15^\circ, 20^\circ, 25^\circ$ and $30^\circ$ resolutions for neutrino zenith angle. The $1\sigma$~CL uncertainties are fitted with cubic splines. As before, the $\chi_{}^2$ distributions are calculated letting $\sin_{}^{2}\theta_{23}^{}$ and $\Delta m_{31}^2$ vary, while $\delta_{\rm CP}^{}$ is let to vary freely within $\delta_{\rm CP}^{} \in [0, 2\pi]$. The figure indicates that for this setup, the value of $\sin_{}^{2} \theta_{23}^{}$ can be measured by about $\Delta(\sin_{}^2 \theta_{23}^{}) \in [3.1\%, 3.6\%]$ accuracy for NO, where the largest estimated uncertainty corresponds to neutrino zenith angle resolution being $30_{}^{\circ}$. These results are obtained for 2 degrees of freedom (dof). In a similar manner, the corresponding effect on $\Delta m_{31}^2$ resolution would be $\Delta(\Delta m_{31}^2) \in [0.04\%, 0.08\%]$ (2 dof) for NO. The results that are obtained for the IO case are comparable to those that are shown for the NO case.

The effect of the neutrino energy resolution in the sub-GeV neutrino energy bins is similarly studied on the determinations of $\sin_{}^{2}\theta_{23}^{}$ and $\Delta m_{31}^2$. In contrast to the impact that is observed for neutrino zenith angle resolution in Figure~\ref{fig:resolutions}, the resolution on the atmospheric neutrino energy for the sub-GeV neutrino energy bins is found to have small effect on the $1\sigma$~CL uncertainties that could be obtained for $\sin_{}^{2}\theta_{23}^{}$ and $\Delta m_{31}^2$. The small impact of the neutrino energy resolution follows from the first oscillation maximum peaking around $E_\nu^{} \sim 6$~GeV. For such energies, the neutrino energies of atmospheric neutrinos correlate strongly with those corresponding to the respective charged leptons. Furthermore, the importance of the first oscillation maximum in the determination of the $\sin_{}^2 \theta_{23}^{}$ and $\Delta m_{31}^2$ values discussed in Section~\ref{sec:theory} implies that both neutrino energy resolution and neutrino zenith angle resolution in the sub-GeV neutrino energy bins have comparatively lower significance for the expected sensitivities to those observables. This is in contrast with the reconstruction capabilities for the multi-GeV bins, where resolutions for neutrino energy and neutrino zenith angle both have great importance.

\begin{figure}[!t]
    \centering
    \includegraphics[width=0.65\linewidth]{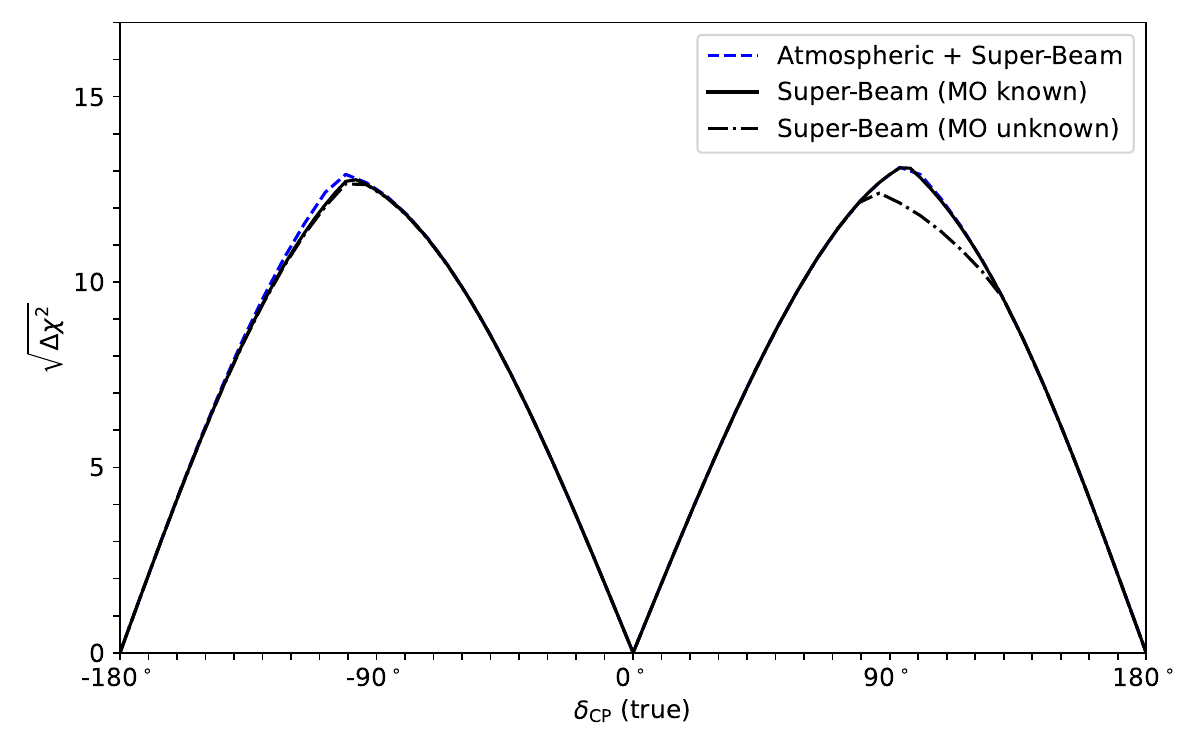}
    \caption{CPV discovery potential in ESSnuSB. The sensitivity to establish CPV is quantified by $\sqrt{\Delta\chi_{}^2}$, which gives the number of standard deviations at which the CP-conserving solution $\sin \delta_{\rm CP}^{} = 0$ can be ruled out for the indicated true value. The sensitivities are shown for the super-beam neutrinos (black dot-dashed curve) and for the combined analysis of super-beam and atmospheric neutrino data (blue dashed curve), assuming that mass ordering is not known. Additionally, the sensitivity is shown for the super-beam data when mass ordering is already known (black solid curve). In all three cases, the true mass ordering is assumed to be NO.}
    \label{fig:CPV-discovery}
\end{figure}

While atmospheric neutrinos are more sensitive in constraining the parameter space for $\sin_{}^2 \theta_{23}^{}$ and $\Delta m_{31}^2$ as a result of the access to the first oscillation maximum, the super-beam neutrino program at ESSnuSB provides the most sensitive probe to CPV. This is exemplified in Figure~\ref{fig:CPV-discovery}, where the sensitivity to exclude the CP-conserving values $\sin_{} \delta_{\rm CP}^{} = 0$ are shown as function of the true values of $\delta_{\rm CP}^{}$. The CPV sensitivity is presented for two separate cases, {\em i.e.}, when atmospheric neutrinos and super-beam neutrinos from ESSnuSB are analysed jointly (blue dashed curve) and when super-beam neutrino samples are studied alone (black dot-dashed curve). It is assumed that neutrino mass ordering is not known. We additionally show the sensitivity to CPV discovery for the super-beam neutrinos assuming that neutrino mass ordering is known (black solid curve). When the super-beam neutrino data is analysed alone, the sensitivity to rule out the CP-conserving values of $\delta_{\rm CP}^{}$ would be about $\sqrt{\Delta \chi_{}^2} \approx 13$ for the maximally CP-violating values true values $\delta_{CP} ^{} = -90^\circ$ and $90^\circ$. Using the Wilks' theorem, this result would be equivalent of $13\sigma$~CL sensitivity with 1 dof. In a similar manner, it can be determined that the $5\sigma$~CL sensitivity limit would be reached for slightly more than $70\%$ of all possible $\delta_{\rm CP}^{}$ values. When the atmospheric neutrino data is included in the analysis, the sensitivity to exclude $\sin \delta_{\rm CP}^{} = 0$ would improve marginally for the true values $\delta_{\rm CP}^{} \in [-120^\circ, -90^\circ]$. For the other possible true values of $\delta_{\rm CP}^{}$, the improvement from atmospheric neutrinos would become negligible for $\sqrt{\Delta \chi_{}^2}$ and the fraction of $\delta_{\rm CP}^{}$ would not increase. This is also in line with the right panel of Figure~\ref{fig:allowed_values}, where no dependence on $\delta_{\rm CP}^{}$ is found for atmospheric neutrinos. As such, including the atmospheric neutrino sample in the numerical analysis would yield no significant advantage in the search for CPV in the case where neutrino mass ordering is already known. On the other hand, if the neutrino mass ordering is not known at the time when ESSnuSB data is analyzed, the sensitivity from the super-beam neutrinos only would decrease for $\delta_{\rm CP}^{} \in [80^\circ, 130^\circ]$. In that case, the loss in the CPV sensitivity loss could be restored by including the atmospheric neutrino sample, as can be seen in the right panel of Figure~\ref{fig:CPV-discovery}. Atmospheric neutrinos are known to be sensitive to neutrino mass ordering and could potentially solve the question on neutrino mass ordering~\cite{ESSnuSB:2024wet}. A combined analysis of atmospheric and super-beam neutrinos would therefore restore the sensitivity for $\delta_{\rm CP}^{} \sim 90^\circ$ even in the case if neutrino mass ordering were assumed to be unknown.

\begin{figure}[!t]
    \centering
    \includegraphics[width=0.65\linewidth]{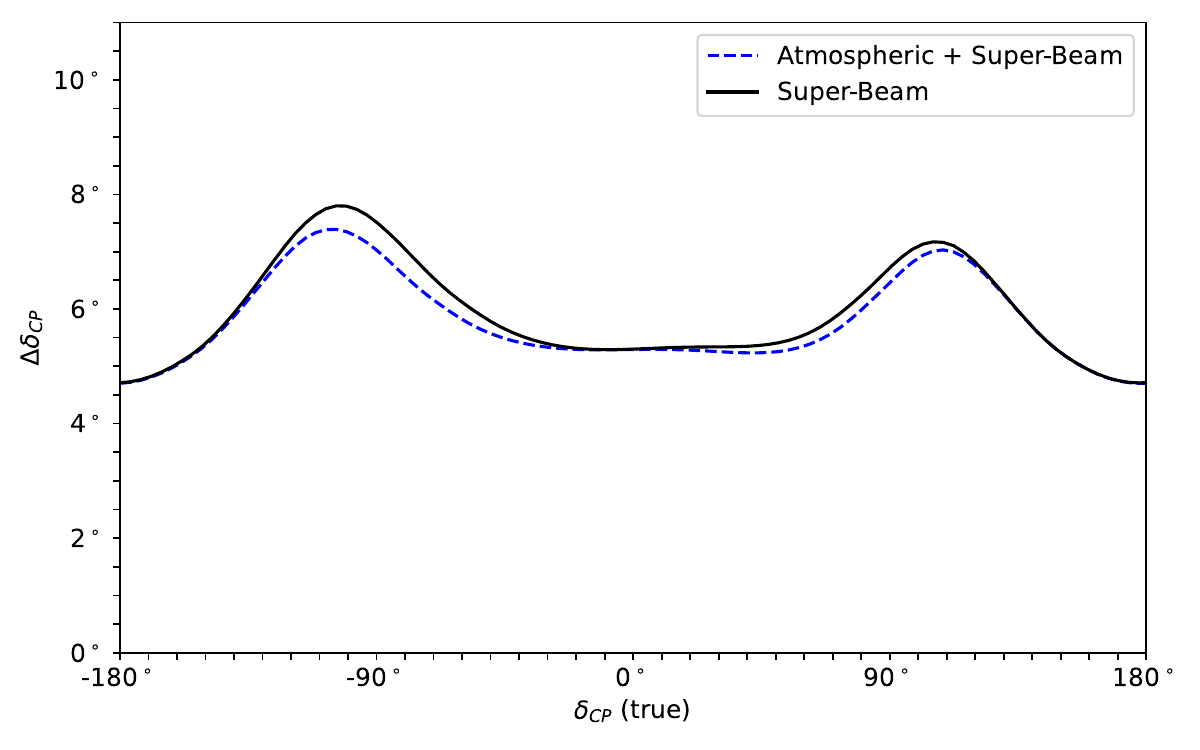}
    \caption{Resolution to measure the leptonic CP phase $\delta_{CP}$ at ESSnuSB. $\Delta \delta_{\rm CP}^{}$ indicates the $1\sigma$~CL resolution at which $\delta_{\rm CP}^{}$ can be measured at ESSnuSB with super-beam neutrinos (black solid curve) and with a combined analysis of super-beam neutrinos and atmospheric neutrinos (blue dashed curve) observed by ESSnuSB FD. The resolution for  measuring $\delta_{\rm CP}^{}$ is shown as a function of true values for $\delta_{\rm CP}^{}$. The true neutrino mass ordering is assumed to be NO. The result is the same regardless of whether or not mass ordering is known prior to the measurement.}
    \label{fig:CPV-resolution}
\end{figure}

The synergy in studying both atmospheric neutrinos and super-beam neutrinos at ESSnuSB is the most evident in measuring the value of $\delta_{\rm CP}^{}$. In Figure~\ref{fig:CPV-resolution}, the resolution at which the value of $\delta_{\rm CP}^{}$ could be determined by $1\sigma$~CL at ESSnuSB is shown as a function of the possible true values of $\delta_{\rm CP}^{}$. The resolutions are shown in the case where both the atmospheric neutrino samples and the super-beam neutrino samples are included in the same analysis (blue dashed curve) and also in the case where the expected super-beam neutrino sample is considered alone (black solid curve). The true neutrino mass ordering is assumed to be NO. Resolution on $\delta_{\rm CP}^{}$ is found to be the same regardless of whether true neutrino mass ordering is known. It can be seen in Figure~\ref{fig:CPV-resolution} that the super-beam neutrino data would be able to determine the value of $\delta_{\rm CP}^{}$ with resolutions $\Delta \delta_{\rm CP}^{} \in [4.7^\circ, 7.8^\circ]$, where the largest uncertainty would be obtained if the true value of $\delta_{\rm CP}^{}$ were near the maximally CP-violating value $\delta_{\rm CP}^{} = -90^\circ$. In comparison, at the recent global-fit value $\delta_{\rm CP}^{} = 212^\circ$, the resolution at which $\delta_{\rm CP}^{}$ could be measured would be about $\Delta \delta_{\rm CP}^{} \approx 5.5^\circ$. When the atmospheric neutrino data is included in the numerical analysis, the resolution for $\delta_{\rm CP}^{}$ would increase for the majority of $\delta_{\rm CP}^{}$ values. The improvement in the $\delta_{\rm CP}^{}$ resolution due to atmospheric neutrinos would be about $0.4^\circ$ at $\delta_{\rm CP}^{} = -90^\circ$, indicating a resolution of $\Delta \delta_{\rm CP}^{} \approx 7.1^\circ$. The most significant improvement in $\Delta \delta_{\rm CP}^{}$ would be $0.5^\circ$. The overall resolution for $\delta_{\rm CP}^{}$ would be $\Delta \delta_{\rm CP}^{} \in [4.7^\circ, 7.5^\circ]$. This improvement would primarily originate from the improved sensitivities to $\sin_{}^2 \theta_{23}^{}$ and $\Delta m_{31}^2$. If the true value of $\delta_{\rm CP}^{}$ were at the current best-fit value $212^\circ$, the contribution from atmospheric neutrinos would be negligible. A similar distribution for $\Delta \delta_{\rm CP}^{}$ can be obtained for the case where the true neutrino mass ordering is IO.

It is found that the effect of including atmospheric neutrino data in the calculation of the CPV sensitivities and the $\delta_{\rm CP}^{}$ resolutions presented in Figures~\ref{fig:CPV-discovery} and~\ref{fig:CPV-resolution}, respectively, can be approximated by including appropriate Gaussian priors for $\sin_{}^2 \theta_{23}^{}$ and $\Delta m_{31}^2$ in $\chi_{\rm priors}^2$. It has been checked that the same results can be obtained both by co-analysing atmospheric and superbeam neutrino data and by replacing atmospheric neutrino data with Gaussian priors. This observation confirms that the perceived improvement in the $\delta_{\rm CP}^{}$ resolution does indeed come from the enhanced constraints on $\sin_{}^2 \theta_{23}^{}$ and $\Delta m_{31}^2$ by the atmospheric neutrino data.

\section{\label{sec:concl}Summary and outlook}

ESSnuSB proposes to use high-powered accelerator-based neutrino beams to measure the leptonic CP phase $\delta_{\rm CP}^{}$ near the second oscillation maximum. With observing neutrinos from its super-beam alone, the ESSnuSB setup would be able to determine the value of $\delta_{\rm CP}^{}$ by $7.8^\circ$ resolution ($1\sigma$~CL) or better. It is shown in this work that including the atmospheric neutrinos data that could be observed in ESSnuSB FD would enhance the resolution on $\delta_{\rm CP}^{}$ by approximately $0.4^\circ$ for $\delta_{\rm CP}^{} = -90^\circ$. The overall resolution is found to be $\Delta \delta_{\rm CP}^{} \in [4.7^\circ, 7.5^\circ]$. The improvement in $\Delta \delta_{\rm CP}^{}$ mainly arises from the enhanced precision on $\sin_{}^2 \theta_{23}^{}$ and $\Delta m_{31}^2$ due to the inclusion of atmospheric neutrinos, which could provide an access to the first oscillation maximum in multi-GeV neutrino energies. It is also shown in this work that the sensitivity to $\sin_{}^2 \theta_{23}^{}$ and $\Delta m_{31}^2$ would not be significantly affected by the event reconstruction capabilities in sub-GeV neutrino energies. Therefore, ESSnuSB would provide the most precise measurement on $\delta_{\rm CP}^{}$.

In addition to the measurement of $\delta_{\rm CP}^{}$, a joint analysis of super-beam neutrinos and atmospheric neutrinos at ESSnuSB would provide an independent measurement on the neutrino mixing parameters. The simultaneous access on neutrino oscillations at the first and second oscillation maxima by the super-beam means that the ESSnuSB setup can be expected to be sensitive to parameters $\theta_{12}^{}, \theta_{23}^{}, \Delta m_{21}^2, \Delta m_{31}^2$ and $\delta_{\rm CP}^{}$. Atmospheric neutrinos can contribute to those measurements further by providing stringent constraints on parameters $\theta_{23}^{}$ and $\Delta m_{31}^2$. It has been shown in this work that the atmospheric neutrino samples that are expected for the ESSnuSB setup would yield more stringent constraints on the values of $\sin_{}^2 \theta_{23}^{}$ and $\Delta m_{31}^2$ than what would be expected for super-beam neutrinos. In contrast, constraints on $\delta_{\rm CP}^{}$ would be derived from the super-beam neutrinos alone, while atmospheric neutrinos would yield no significant contribution. Therefore, the sensitivities to $\sin_{}^2 \theta_{23}^{}$ and $\Delta m_{31}^2$ by atmospheric neutrinos would be highly complementary to super-beam neutrinos.

In conclusion, it has been shown in this work that atmospheric neutrinos can provide complementary information for the long-baseline neutrino program at ESSnuSB. The large statistics that are expected for atmospheric neutrinos at ESSnuSB FD would not only provide notable sensitivities to neutrino oscillation parameters $\theta_{23}^{}$ and $\Delta m_{31}^2$, but the promise of the super-beam program would also allow to use atmospheric neutrinos from ESSnuSB FD to provide better sensitivities to probe physics beyond the Standard Model.

\section*{Acknowledgments}
Funded by the European Union, Project 101094628. Views and opinions expressed are however those of the author(s) only and do not necessarily reflect those of the European Union. Neither the European Union nor the granting authority can be held responsible for them. We acknowledge further support provided by the following research funding agencies: Centre National de la Recherche Scientifique, France; Deutsche Forschungsgemeinschaft Projektnummer 423761110, and under the Excellence Strategy of the Federal Government and the L{\"a}nder, Germany; Ministry of Science and Education of Republic of Croatia grant No. PK.1.1.10.0002; the European Union’s Horizon 2020 research and innovation programme under the Marie Sk{\l}odowska-Curie grant agreement No 860881-HIDDeN; the European Union NextGenerationEU, through the National Recovery and Resilience Plan of the Republic of Bulgaria, project No. BG-RRP-2.004-0008-C01; Roland Gustafssons Stiftelse f\"or teoretisk fysik, Sweden; Swiss National Science Foundation (SNSF), Croatian Science Foundation (HRZZ) and National Recovery and Resilience Plan (NPOO) via the grants MAPS IZ11Z0$\_$230193 and DOK-NPOO-2023-10-1262; as well as support provided by the universities and laboratories to which the authors of this report are affiliated, see the author list on the first page. The authors are grateful for the computing resources that were provided by the Division of Condensed Matter Theory at KTH Royal Institute of Technology.
\bibliographystyle{apsrev4-2}
\bibliography{references}
\end{document}